\documentclass[10pt,letterpaper]{article}
\usepackage[top=0.85in,left=2.75in,footskip=0.75in,marginparwidth=2in]{geometry}

\usepackage[utf8]{inputenc}

\usepackage[numbers,sort&compress]{natbib}

\usepackage{nameref}
\usepackage{xcolor}
\usepackage{hyperref}
\hypersetup{colorlinks,allcolors=blue!50!black}

\usepackage[right]{lineno}

\usepackage{microtype}
\DisableLigatures[f]{encoding = *, family = * }

\raggedright
\usepackage{changepage}

\usepackage[aboveskip=1pt,labelfont=bf,labelsep=period,singlelinecheck=off]{caption}

\makeatletter
\renewcommand{\@biblabel}[1]{\quad#1.}
\makeatother

\usepackage{lastpage,fancyhdr,graphicx}
\usepackage{epstopdf}
\fancyheadoffset[L]{2.25in}
\fancyfootoffset[L]{2.25in}

\usepackage{color}

\definecolor{Gray}{gray}{.25}

\usepackage{graphicx}

\usepackage{sidecap}

\usepackage{wrapfig}
\usepackage[pscoord]{eso-pic}
\usepackage[fulladjust]{marginnote}
\reversemarginpar

\usepackage{amsmath}
\usepackage{amsfonts}
\usepackage[skip=8pt]{parskip}
\usepackage{subfig}
\usepackage{fontawesome}

\begin{document}
\vspace*{0.35in}

\begin{flushleft}
{\Large
\textbf\newline{Social coordination perpetuates stereotypic expectations and behaviors across generations in deep multi-agent reinforcement learning}
}
\newline
\\
Rebekah A. Gelpí\textsuperscript{1,2,3,*},
Yikai Tang\textsuperscript{1,2},\\
Ethan C. Jackson\textsuperscript{1,2},
William A. Cunningham\textsuperscript{1,2,3,4}
\\
\bigskip
\textbf{1} Department of Psychology, University of Toronto
\hspace{1em}
\textbf{2} Vector Institute
\\
\textbf{3} Schwartz Reisman Insitute for Technology and Society, University of Toronto
\\
\textbf{4} Department of Computer Science, University of Toronto
\\
\bigskip
* To whom correspondence should be addressed: \href{mailto:rebekah.gelpi@mail.utoronto.ca}{rebekah.gelpi@mail.utoronto.ca}
\\
\bigskip
Preprint. Under review.

\end{flushleft}

\section*{Abstract}
Despite often being perceived as morally objectionable, stereotypes are a common feature of social groups, a phenomenon that has often been attributed to biased motivations or limits on the ability to process information. We argue that one reason for this continued prevalence is that pre-existing expectations about how others will behave, in the context of social coordination, can change the behaviors of one’s social partners, creating the very stereotype one expected to see, even in the absence of other potential sources of stereotyping. We use a computational model of dynamic social coordination to illustrate how this “feedback loop” can emerge, engendering and entrenching stereotypic behavior, and then show that human behavior on the task generates a comparable feedback loop. Notably, people’s choices on the task are not related to social dominance or system justification, suggesting biased motivations are not necessary to maintain these stereotypes.

\section*{Introduction}

The widespread presence of stereotyping—generalized, and often exaggerated, inferences about individuals based on their social category membership—is a broad social concern that presents a challenge to group diversity. Groups contain individuals who may have differing characteristics, skills, and needs; stereotypes threaten this by projecting similarity onto members of a social category that may not accurately reflect the diversity and variability of individuals within it. Many classic and contemporary theories of social cognition have sought to explain why otherwise well-meaning individuals might nevertheless engage in stereotyping. Some of these theories suggest that motivations such as promoting one’s group at the expense of another’s \citep{brewer_psychology_1999, austin_integrative_1979} or justifying existing inequalities \citep{jost_decade_2004, jost_role_1994, sidanius_social_2001} conflict with norms against stereotyping; others argue that the limits of human cognition make overly simplistic generalization inevitable \citep{fiske_continuum_1990, macrae_stereotypes_1994, sherman1998stereotype}. 

We argue that part of the reason stereotyping remains so prevalent despite the existence of norms against is because these stereotypes also result from the dynamics of social coordination, where social partners shift their behaviours to align with the pre-existing expectations of the social environment in which they find themselves. In large societies, people must interact with strangers with whom they have almost no experience, requiring them to “go beyond the information given” \citep{bruner1957going} to make inferences about the traits, motivations, and beliefs of others based on secondary cues. As a result, a feedback cycle emerges in which the target of an expectation who successfully coordinates with a social partner amplifies that partner's expectation that the target will continue to behave according to that expectation.

Using a combination of computational and behavioral work, we illustrate how such dynamics can emerge from the cognitive processes of individuals operating within a social system, and how agents in these systems can inadvertently contribute to a social structure that helps create, maintain, and strengthen social biases.

\subsection*{Statistical Regularities and Social Conventions}

Modern human social groups are large and deeply complex. In order to successfully develop and maintain these societies, people must successfully learn to coordinate with one another, dividing labour (whether physical, social, or cognitive) between individuals to allow for specialization and large-scale cooperation \citep[e.g.,][]{binmore2005natural, henrich2008division, oconnor_origins_2019, sloman2018knowledge}.

Coordinating successfully in such large groups poses a challenge: reliably knowing how to assign tasks with an individual requires understanding that individual's skill set, but as the size of a social group increases, representing each individual within it with high fidelity likewise incurs a greater cognitive cost. Beyond a certain point, this poses a computationally intractable problem, as finite cognitive resources constrain the number of simultaneous close social relationships we are able to entertain \citep{dunbar_social_1998, mac_carron_calling_2016, tamarit_beyond_2022, tamarit_cognitive_2018}. Several theories of social cognition have therefore characterized people as “efficiency experts” \citep[e.g.,][]{fiske_continuum_1990, macrae_stereotypes_1994, sherman1998stereotype}) for whom group-based generalizations result from an attempt to leverage the utility of low-cost category knowledge, thus avoiding the more effortful task of individuation. Generalizing these regularities to unfamiliar others, in turn, facilitates the development of shared knowledge and the ability to coordinate one’s behavior.

Given these potential advantages, social norms or conventions that members of a group ought to engage in a particular behavior or share a certain belief can further simplify the space of expected behavior, easing coordination by making each group member more predictable \citep{wheeler_ideology_2020}. For example, computational accounts of the development and maintenance of conventions such as communication and cooperative foraging have emphasized the need not only to coordinate with individuals but to abstract these rules for interactions away from individual relationships and generalize them to a larger community \citep[e.g.,][]{hawkins_partners_2022}, supported by expectations and punishments for failing to align one's behavior with the coordinated norm \citep[e.g.,][]{burke_social_2011, fehr_third-party_2004, koster_spurious_2022, vinitsky_learning_2022}.

The codification of regularities into expectations and social conventions, while often providing substantial benefits to group members, can come with potential drawbacks, both to individual members within the group and also to the group as a whole. By compressing our highly complex social environments into easily interpretable generalized beliefs about social categories, individuals whose skills or preference do not match what a convention dictates can be unfairly disadvantaged. 

Game theory has long used economic games such as the Bach-Stravinsky game \citep[also known as the Battle of the Sexes; see e.g.,][]{luce1957games, osborne1994course} to model how self-interested agents (and subgroups of agents) develop stable strategies for coordinating with one another. Coordination in such games reliably outperforms uncoordinated behavior, but in many cases, the rewards of coordination are not equally distributed. At the individual level, even when groups have an equal underlying distribution of skills (i.e., which group specializes in which task is ultimately arbitrary), the fact that many specialized tasks within human societies require substantial investment creates strong incentives for individuals to coordinate, even when it is personally detrimental \citep[e.g.,][]{hadfield1999coordination}. Coordination can also result in disadvantages for groups as a whole; for example, when coordination involves tasks with unequal rewards, one group is reliably disadvantaged \citep{oconnor_origins_2019}, particularly when tasks provide surpluses \citep{henrich2008division} or when one group is smaller than another \citep{bruner_minority_2019, oconnor_emergence_2019}. In all of these cases, while coordination may be useful in the sense that it results in better outcomes as a whole, it can come at the cost of disadvantaging those whose skills are not reflected by this simplified representation.

\subsection*{Learning Signals and Feedback Loops}

Although game-theoretic analyses have provided a powerful tool for understanding how inequality can emerge from the need to coordinate, these models often do not capture the process by which individuals within a society learn and change their behavior. When learning agents attempt to coordinate, their actions also change one another's reasoning about optimal actions. For example, believing that one's partner prefers to clean and does not like to cook might make one more likely to offer to cook than to clean, all else being equal; however, this also means that one's partner is more likely to offer to clean than to cook in the future, even if the partner actually had no strong feelings to begin with. 

In this way, learning systems can create positive feedback loops between expectations and outcomes. When these inferences involve generalized beliefs about social categories as a whole, they regularize one's experience of social groups more than would be expected by chance, strengthening the associations between a social category and an expectation not just for oneself, but also for agents who are newly introduced to the system.

Another limitation to modelling coordination with game theory is that it is unclear what agents' beliefs about the nature of the regularities being learned are—whether they represent true underlying differences about a group, or whether they are simply an arbitrary convention agreed upon for the sake of efficiency or mutual benefit. If agents believe the outcomes of a coordinated system reflect an essential difference between groups, they may resist changes to that system and justify its resulting inequalities; however, this tendency may be mitigated if the convention is perceived as an arbitrary rule that it is possible to arbitrarily change.

One piece of evidence that agents' pre-existing beliefs can contribute to the feedback loop described above comes from the phenomenon of “behavioral confirmation” observed for a number of real-world social categories, such as race and gender \citep{klein_stereotypes_2003, madon_self-fulfilling_2011}. In one such task by Skrypnek \& Snyder (\citep{skrypnek_self-perpetuating_1982}), male and female students were paired together and tasked with negotiating a division of labor to solve a set of stereotypically masculine and feminine tasks without face-to-face interaction. Male students allocated more masculine tasks to their partners when they believed they were negotiating with a male student; on the following task, the female students who had been perceived as male allocated more masculine tasks for themselves. Thus, the initial perceiver’s stereotypic beliefs about men and women were not only confirmed by the target’s behavior in the first task, but strengthened by the target’s own subsequent choices. Therefore, if stereotypes can emerge from behavioral confirmation, then generalized conventions that disadvantage or misrepresent a group of individuals can emerge even among unbiased individuals with sufficient cognitive capacity to fully represent the complexity of their social interactions. 

This possibility offers a potential answer to a fundamental question in social psychology: why people engage in stereotyping, despite endorsing moral norms against the use of stereotyping. In such situations, people may believe that they are simply drawing statistically appropriate conclusions about how groups are likely to behave \citep{cao_people_2019, vrantsidis_stereotypes_nodate}, even though their own professed belief may itself create, entrench, and strengthen the stereotype that individuals within a group will engage in a particular behavior.

We illustrate the development of stereotypic conventions within populations of skilled agents that learn and change in response to their environment using multi-agent deep reinforcement learning (MARL). MARL models are an emerging tool in the fields of artificial intelligence and cognitive science to understand social interaction and coordination problems \citep{chen_social_2020, jaques_social_2019, lerer_learning_2019, leibo_multi-agent_2017, zhang_bi-level_2020}, particularly those involving the emergence of social conventions \citep{koster_model-free_2020, koster_spurious_2022, vinitsky_learning_2022}, and open several new lines of inquiry that have not been possible using traditional game-theoretic approaches. Using our models, we test two hypotheses regarding the emergence of stereotypic conventions: (1) that these effects are most prominent in larger populations, as coordination around group labels is learned more quickly than the skills of individual agents, and (2) that learning agents develop a structure that is learned and perpetuated by subsequent generations of new agents, such that agents' initial expectations continue to affect the structure of the world even when those agents no longer exist with an environment.

These models allow us to test whether agents optimizing their own anticipated rewards will develop and persist in stereotypic patterns of behavior. However, while these models provide a baseline for what maximizes an agent's expected utility, human beings can act against their economic interests to behave prosocially \citep[e.g.,][]{fehr2003nature, gintis2003explaining, henrich2001search}. Thus, people who are aware that an underlying convention is arbitrary in ways that disadvantage a group would be less likely to endorse that convention—but only to the extent that they are aware such a difference exists. By testing human participants on a comparable scenario to our MARL models, we can directly prompt explicit beliefs about the structure of the environment that they are learning in, allowing us to link the stereotypic actions that may result from social coordination, to the stereotypic beliefs that exist in our daily life. If human participants believe that the conventions observed in our tasks reflect a true, underlying essential difference, then participants on our task should not only make stereotype-consistent choices, they should do so irrespective of individual differences in tolerance for social inequality.

\subsection*{Task Setup}

\begin{wrapfigure}[15]{r}{75mm}
\vspace{-44pt}
\includegraphics[width=75mm]{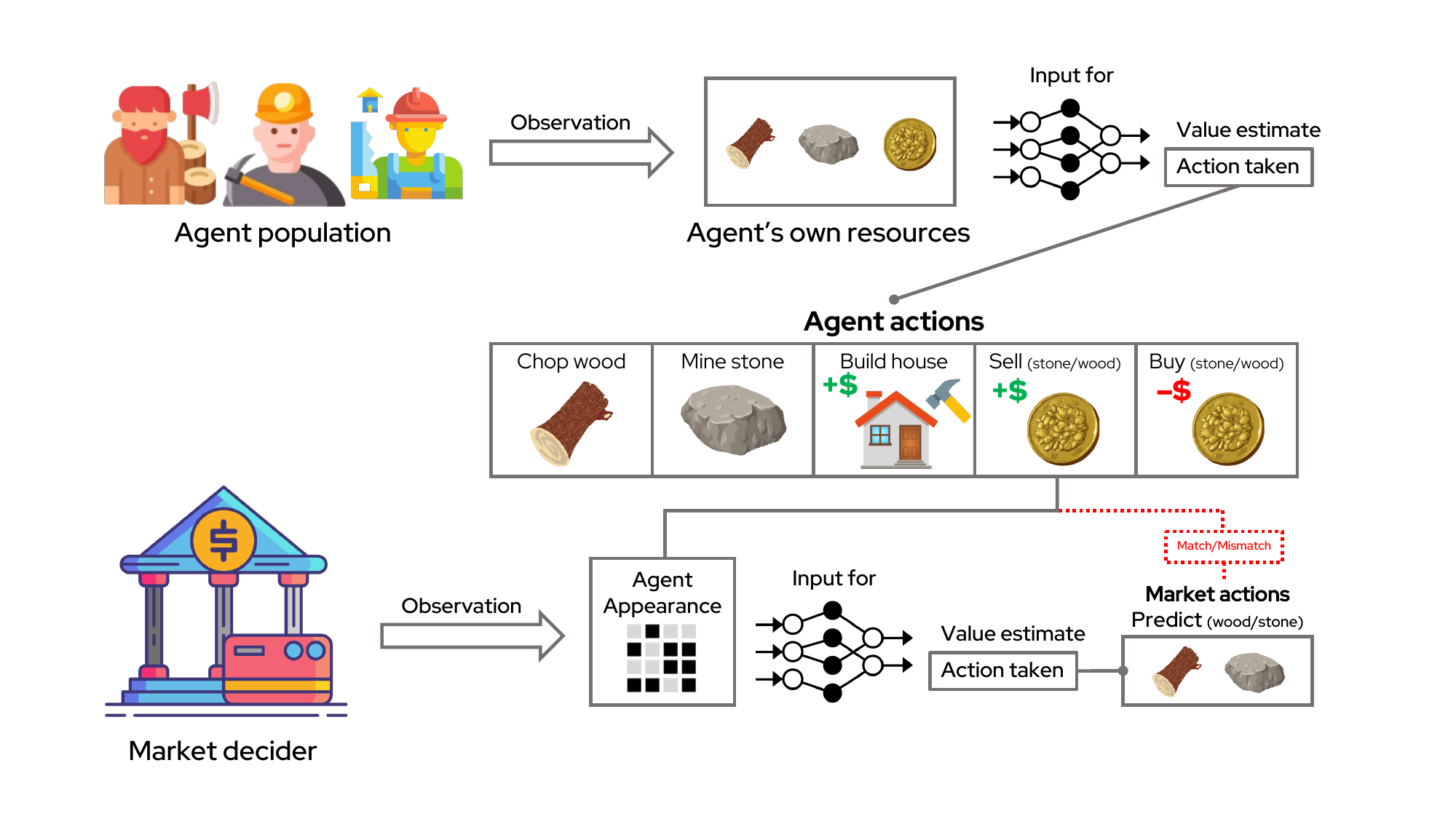}
\caption{\footnotesize{\textbf{Non-technical schematic of task design.} Agents (top) with varying skill specializations observe their own resources and input this information into a neural network to determine their actions. The market decider (bottom) observes the agent's appearance when it chooses to sell to the market, and input this information into its own neural network to determine whether to predict that the agent will sell wood or stone. Successful sales (and the agents'/market decider's reward) depend on the prediction matching the agent's action.}} % add dummy caption - otherwise \label won't work and figure numbering will not count up
\label{fig:fig1} % use \ref{fig1} to reference to this figure
\end{wrapfigure} % avoid blank space here

At a high level, we investigate the effect of this feedback loop on an environment in which there are differences in the benefits or needs of individual members within a social group. For example, a social group might have a larger number of people who are skilled lumberjacks, and a smaller number who are skilled miners. In a large enough social group, this group might be represented as “lumberjacks”, establishing a norm that members of the group ought to chop wood. If the group's success depends on successful coordination, the mechanisms of expectation and enforcement— the “carrot” and the “stick” of convention formation—systematically disadvantage those who would be better served by mining, even as the majority is advantaged. Worse yet, if the disadvantaged miners do accede to the norm, engaging in chopping wood despite their lack of skill, this strengthens the perceived regularity that members of the group are lumberjacks, and worsens the misrepresentation by the broader population of the underlying diversity of the social group.

We adapt the AI economist paradigm \citep{tang_unequal_2023, zheng_ai_2020} to model this environment. Agents in our studies participate in a produce-and-trade task where they collect resources from the environment, which they can use to trade with a market decider (described below) in exchange for rewards (Fig.~\ref{fig:fig1}). By making agents' rewards contingent on successful trades with an external market, this task allows for analyzing how the agents' and markets' simultaneous learning of the dynamics leads to patterns of social coordination. 

Although the coordination demand between the agent and the market may seem to be trivially resolved, e.g. by communicating one's intentions in advance, thus preventing failures due to misaligned behaviors, this setup is identical to a typical coordination problem, and serves as an abstraction for situations where individuals actively select interactions with others about whom they know very little beyond pre-existing beliefs or stereotypes.

Each agent’s input consists of its current reward value as well as its resource total (wood and stone). Each agent takes a single action per turn, with an action space of seven total possible actions. Three actions are independent actions: chopping wood (receive 1 wood), mining stone (receive 1 stone), or building a house (consume 1 wood and 1 stone, receive 15 points of reward). These actions do not involve the market decider, and their success rate depends on the skill levels of each agent (e.g., an agent with a skill level of 0.95 in chopping would successfully receive 1 wood on 95\% of trials spent chopping).

The other four actions are social actions: buying wood, selling wood, buying stone, or selling stone. Successful selling actions require coordination between the agents and the market decider. Whenever an agent decides to sell a resource, the market decider has to predict the resource that the agent is selling; the transaction is only completed if the market decider makes the correct guess about what the agent is attempting to sell. After a successful transaction, the agent trades away 2 units of wood or stone, and receives 1 point of reward. Buying actions do not require coordination; if an agent buys a resource from the market decider, it loses 2 points of reward and receives 1 unit of wood or stone.

Unlike the agents, the market decider does not necessarily act at every timestep, nor is it limited to acting only once every timestep; rather, it only takes an action when agents opt to sell to the market decider. In each trade, the market decider’s input consists of a unique 16-digit binary code corresponding to the agent’s identity. The market decider receives positive or negative rewards depending on its prediction of agents’ selling behaviors. It receives 1 point of reward when an agent successfully sells a resource to it, loses 1 point of reward if it does not correctly predict the agent’s behavior, and loses 0.3 points of reward if it makes a successful prediction, but the agent does not have enough resources to make a successful transaction.

\section*{Results}

The design of the social group structure in our studies is aimed to enable the possibility of the emergence of social conventions. To this end, we focus on the behaviors of mining and chopping specialist agents, and the degree to which they adhere to conventions in groups that are composed of a majority of choppers or miners. When selling to the market decider, these agents can behave according to their skill (by offering to sell the resource that they are skilled at collecting), or contrary to their skill (by offering to sell the resource that they are not skilled at collecting). Below, we examine the proportion of skill-based predictions by the market decider, and the proportion of skilled and unskilled behaviors by agents to measure the emergence of group-level conventions among the groups.

\subsection*{Study 1: Individuation}

To determine whether our model has the capacity to fully individuate all of the agents, we first conduct a series of baseline experiments, in which agents are assigned their identity codes randomly, and there is no correlation between these codes and the agent’s skill. Thus, for the agents and the market decider to successfully coordinate, the market decider must learn to predict each agent’s skill based on its unique identity code. To the extent that these models can successfully learn to predict agents’ behavior according to their skills, this indicates that any stereotyping effects observed in later studies are not the product of a model 
 that is incapable of learning the complexity of a large social environment. Further, we can observe how differences in population size are reflected in the time it takes our models to learn individuals’ skills.

Our results from this study suggest that the basic structure of the task is learnable by both the agents and the market decider, giving us a benchmark of how well our market decider model is capable of representing individuals’ unique skill. Agents predominantly behave according to their skill; for example, skilled choppers mostly chop wood and sell wood, while skilled miners mostly mine stone and sell stone (Fig.~\ref{fig:fig2}A-B). Meanwhile, the market decider learns to expect that agents will behave according to their skilled behavior to a high degree of accuracy; the market decider’s predictions match agents’ skilled behavior 95\% of the time (Fig.~\ref{fig:fig2}C). 

The market decider is most accurate at making skill-based predictions when the size of the population is smaller, $F(3,7988) = 1692.53$, $p < .001, \eta^{2}_{p} = 0.39$, and the market decider's predictions become more accurate over time, $F(1,7988) = 14127.42$, $p < .001, \eta^{2}_{p} = 0.64$. However, the larger the population, the longer it takes for the market decider to reach a high level of accuracy, $F(3,7988) = 615.39$, $p < .001, \eta^{2}_{p} = 0.19$ (Fig.~\ref{fig:fig3}A). Broadly, we find that although the time taken to learn the population varies based on the group size, the market decider is able to predict how agents are likely to behave according to their true underlying skills. Due to the structure of the learning task in this condition, however, the market decider’s accuracy in its predictions about how unobserved agents are likely to behave cannot change from 50\% throughout training, because the skills and identity of unobserved agents is uncorrelated with the skills and identity of existing agents.

\subsection*{Study 2: Regularity}

As population sizes increase, people have fewer opportunities to learn about each individual within a population; correspondingly, people may rely on secondary cues such as group identity if these cues can predict any existing variance. To the extent that these secondary cues are used for prediction in coordination tasks, and these predictions lead to expectations that influence others’ behavior, we should expect that when a secondary cue is statistically correlated with a skilled behavior, all agents with that cue should behave more conventionally, even if this means behaving in a way that conflicts with the agent’s personal skill.

\vspace{.5cm}
\begin{adjustwidth}{-2in}{0in}
\begin{flushright}
\includegraphics[width=180mm]{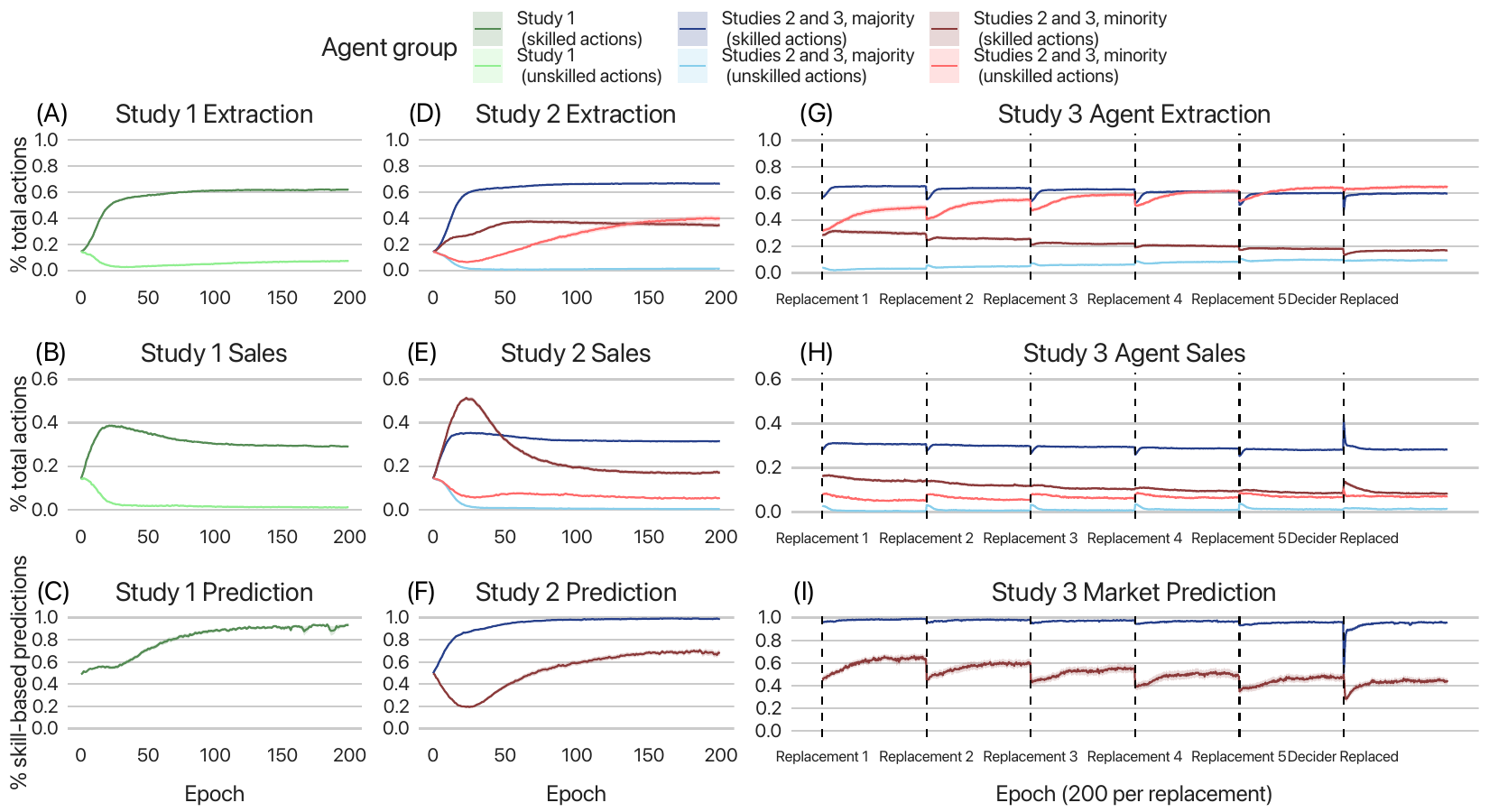}
\end{flushright}
\justify 
\color{Gray}
\textbf{Figure 2. Agent and market behavior for Studies 1-3.} Agent extraction and selling behavior, and market decider’s predictions, for Studies 1–3, with a population size of 300 agents. (A, B) Agents' extraction (A) and selling (B) behavior in Study 1. Agents learn to extract and sell resources that match their personal skills (dark green), and rarely extract or sell resources that they are unskilled at collecting (light green). (C) Market decider's predictions in Study 1. The market decider's predictions match the agents' personal skills 95\% of the time. (D, E) Agents' extraction (D) and selling (E) behavior in Study 2. Majority agents almost exclusively match the stereotypic belief about the group as a whole, extracting and sell resources that they are skilled at collecting (dark blue) rather than the resources that they are unskilled at collecting (light blue). Minority agents, on the other hand, progressively extract and sell lower proportions of the resources they are skilled at collecting (dark red) and higher proportions of those they are unskilled at collecting (light red) over time. (F) Market decider's predictions in Study 2. The market decider almost always predicts that majority agents (dark blue) will act according to their skill, but predicts that minority agents will often not act according to their skill (dark red). (G, H) Agents' extraction (G) and selling (H) behavior in Study 3. As agents are iteratively replaced (dashed lines), the patterns observed in Study 2 strengthen; majority agents continue to mostly extract and sell resources they are skilled at collecting (dark blue) rather than unskilled at collecting (light blue). Minority agents' behavior becomes more stereotypic, mostly extracting the resources they are unskilled at collecting (dark red) rather than skilled at collecting (light red), and selling both skilled and unskilled resources at comparable rates. (I) Market decider's predictions in Study 3. As new agents are introduced (dashed lines), the market decider's predictions for the minority agents (dark red) becomes progressively less skill-based and more stereotypic, while its predictions for the majority agents (dark blue) are almost exclusively skill-based.
\end{adjustwidth}
\begin{figure}
    \captionsetup{labelformat=empty}
    \caption{}\label{fig:fig2}
\end{figure}

In Study 2, we examine this by making each agent’s skills statistically correlated with three of the digits of the agent’s identity code, thus functioning as a group label. While 50\% (the majority) of the agents with a given group label share a skill specialization, the remaining 50\% is split evenly between the other two skill specializations (the minorities). In the groups where the majority of agents are chopping or mining specialists, this yields a ratio of 2:1 between the size of the chopping or mining specialist majority and the corresponding mining or chopping specialist minority, respectively. 

Critically, all individuating information from Study 1 is still present, so it is possible for the model to fully learn each agent’s skilled behavior. However, because both the market decider and the agents are learning the task dynamics simultaneously, if the market decider develops expectations about agents based on the group label before it has learned individual agents’ skills, then agents may begin to display behavioral confirmation, establishing a stereotypic convention that regularizes the market decider’s learning signal and precludes any ability to learn about individuals’ unique skilled behaviors.

We find this behavioral confirmation effect emerges in Study 2. The market decider quickly learns to predict that agents with a given group label will mostly attempt to sell the same resource that the majority of their group is skilled at collecting, i.e. predicting that most agents in the majority-miner group will bring stone, and most agents in the majority-chopper group will bring wood (Fig.~\ref{fig:fig2}F). As a result, agents in these conditions act according to the group label observed by the market decider, rather than their own personal skills: the minority group engages in a significantly lower proportion of skilled actions than agents in Study 1 (extraction: $B = -0.209$,  $SE = .002$, $95\% \ CI = [-0.213, -0.205]$, $z = -122.66$, $p < .001$, $\text{Cohen's} \ d = -1.42$; sales: $B = -0.11$, $SE = 0.001$, $95\% \ CI = [-0.111, -0.109]$, $z = -83.02$, $p < .001$, $\text{Cohen's} \ d = -0.96$; Fig.~\ref{fig:fig2}D--E), and by contrast, majority group members engage in a slightly greater proportion of skilled actions than agents in Study 1 (extraction: $B = 0.065$, $SE = 0.002$, $95\% \ CI = [0.061, 0.069]$, $z = 38.15$, $p < .001$, $\text{Cohen's} \ d = 0.44$; sales: $B = 0.037$, $SE = 0.001$, $95\% \ CI = [0.035, 0.041]$, $z = 27.90$, $p < .001$, $\text{Cohen's} \ d = 0.32$). Thus, even though the same market decider model was capable of learning individuals’ skills to a high degree of accuracy, the confounded statistical cue leads the model to engage in stereotyped predictions.

\begin{figure}[ht]

\includegraphics[width=\textwidth]{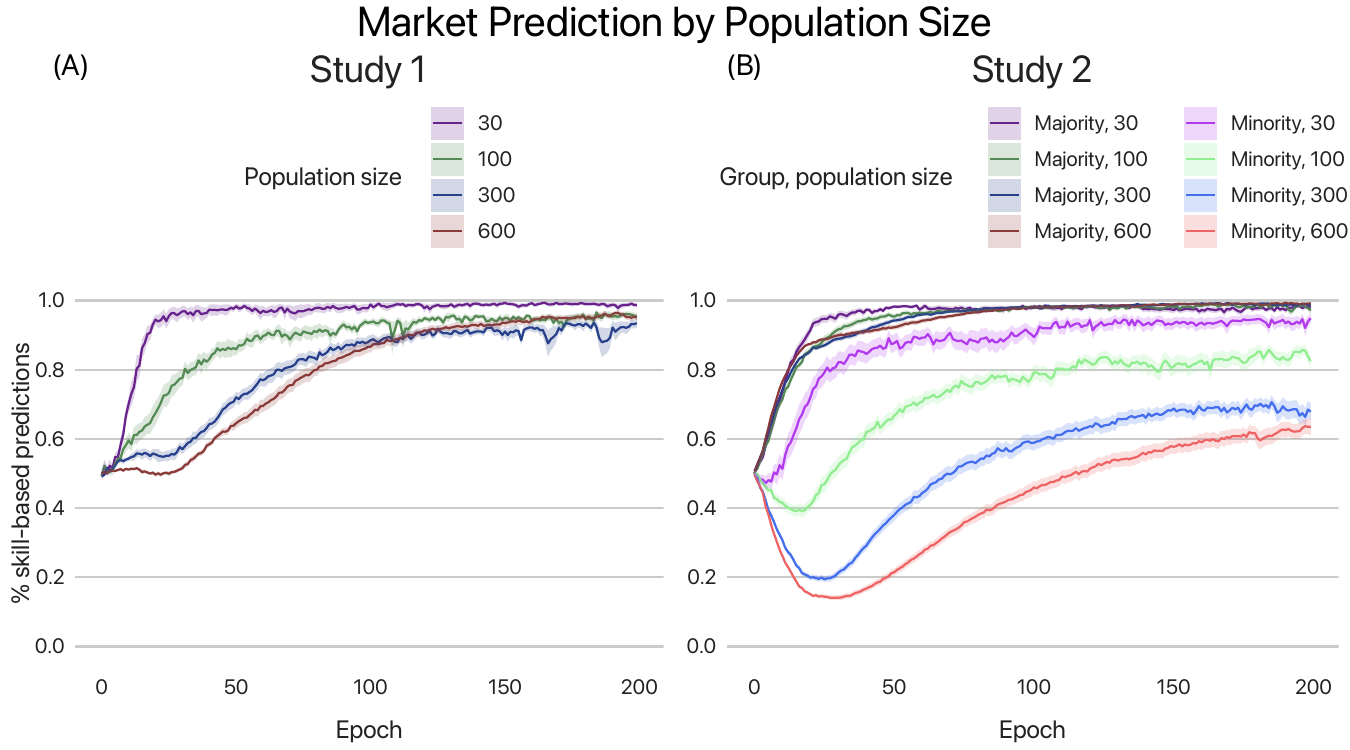}

\caption{\textbf{Market predictions with differing population sizes.} In Study 1 (A), the market decider learns to predict that agents will sell the resource that matches their skill. Larger population sizes take longer to learn, but does not affect the ultimate frequency of skill-based predictions. In Study 2 (B), the market decider’s predictions quickly converge to skill-based predictions the majority group members (i.e., chopping specialists sell wood and mining specialists sell stone); however, the market decider’s predictions for minority members depends on group size, such that the market decider makes mostly skilled-based predictions for the minority in the population size of 100, but almost exclusively stereotypic predictions for the population size of 600.}

\label{fig:fig3}

\end{figure}

Study 2 also illustrates how the complexity of learning each individual as population size increases affects the market decider’s predictions when there is a statistical regularity. It is more difficult and time-consuming to learn the individual skill of each agent in a larger population. On the other hand, a statistical regularity—that more agents with a given group label have a particular skill—can be learned more quickly because of its simplicity. Further, it does not become more difficult to learn as the population size increases. In a population size of 30, the market decider predicts that agents with the minority skill in a group will behave according to their own skill most of the time, albeit still less often than the majority agents ($B = 0.080$, $SE = 0.002$, $95\% \ CI = [0.076, 0.085]$, $z = 43.01$, $p < .001$, $\text{Cohen's} \ d = 0.78$); however, as the population size increases, the market decider’s predictions, and consequently the agents’ behaviors, become more highly stereotyped, with minority agents becoming much less likely to behave according to their skills than majority agents (all $z \geq 116.99$, $p < .001$, Cohen's $d \geq 2.14$; Fig.~\ref{fig:fig3}B). 

As both the market decider and the agents are learning the task dynamics simultaneously, the fact that the market decider rapidly learns and applies a statistical cue before it is able to learn individuals’ skilled behavior means that agents begin to change their own behaviors to match the market decider’s predictions. In other words, these predictions result in a behavioral confirmation effect—the market decider’s predictions lead agents to behave in a more stereotypic fashion. This behavior is, in turn, observed by the market decider, strengthening and entrenching the prediction it makes about these agents.

\subsection*{Study 3: Generational Transmission}

In Study 2, we demonstrate that behavioral confirmation can develop simply by including a statistical cue that is more easily learned than an individual’s skill. This effect, in particular, disadvantages agents whose skill specialization is a minority within the larger group.

However, social environments are not static, and it is possible for underlying characteristics such as the correlation between a secondary cue and a skill to change over time. Stereotypic expectations developed over the course of history might be capable of maintaining stereotypic behaviors long after any underlying group differences have faded. In Study 3, we examine how stereotype-consistent behavior can be maintained from stereotypic group-based expectations, even when groups have no underlying differences.

In this study, we continue to train the models from Study 2, but begin to iteratively replace agents from the population, 20\% at a time, with agents whose skill specializations are uncorrelated with the group label (i.e., an equal proportion of chopping and mining specialists). After all agents have been replaced, we additionally replace the market decider with a new, untrained market decider. At this point, there is no longer any statistical regularity between the group label and the agents’ skills, similar to Study 1, and the new market decider has not developed expectations about any of the agents.

\begin{figure}[ht]

\includegraphics[width=\textwidth]{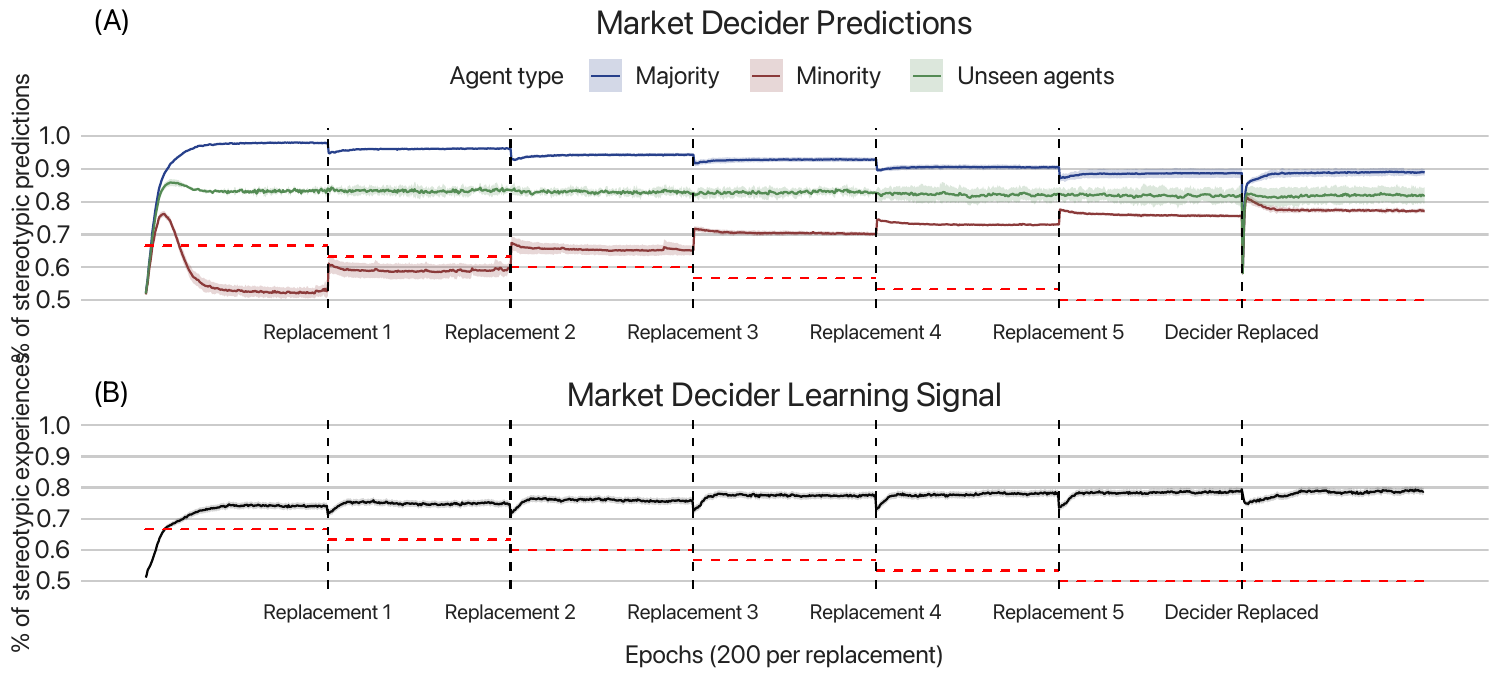}

\caption{\textbf{Raw learning signal and market predictions for unobserved agents.} (A): Proportion of stereotypic predictions by the market decider for previously observed majority agents (blue), minority agents (red), and held-out unobserved agents (green) in Study 3. The market stably predicts that unseen agents will mostly engage in stereotypic actions at a greater rate than the ground truth skill distribution (dashed red line), even as the underlying skill distribution changes. (B): Market decider's raw learning signal for mining and chopping groups. As the market decider's expectations prompt minority agents to engage in against-skill behaviors, the decider's learning signal exaggerates the underlying difference in skill sets, and maintains it after the underlying difference disappears.}

\label{fig:fig4} % \label works only AFTER \caption within figure environment

\end{figure}

Because the initial market decider’s expectations have already been established and reinforced, new agents face a strong pressure to behave according to the existing convention. Regardless of their true skills, new agents are rewarded if they can successfully coordinate with the market decider. As a result, although new agents introduced into the environment are evenly divided between chopping and mining specialists, these agents increasingly behave according to the stereotypic convention, and the difference between majority and minority agents' selling and extracting behavior becomes progressively more stereotyped between the initial replacement and the final replacement, all $B \geq 0.21$, $ SE = 0.002$, $z \geq 109.02$, $p < .001$, $\text{Cohen's} \ d \geq 1.99$ (Fig.~\ref{fig:fig2}G--H). This pattern of increasingly stereotyped behavior by both minority and majority agents suggests that the behavioral confirmation of the market decider can create self-sustaining stereotypic behavior in the absence of underlying group differences. As the market decider's learning signal becomes progressively more stereotypic (Fig.~\ref{fig:fig4}B), the market decider, in turn, develops progressively more stereotypic expectations about the groups as agents are replaced, $B = 0.084$, $SE = 0.003$, $95\% \ CI = [0.078, 0.089]$, $z = 29.59$, $\text{Cohen's} \ d = 1.08$, even though the underlying differences are actually becoming smaller with each replacement (Fig.~\ref{fig:fig2}I, Fig.~\ref{fig:fig4}A).

Lastly, we find that upon replacing the market decider, the agents continue to engage in a pattern of stereotypic behavior. Due to the expectations of the previous market decider, and the convention that emerges as a result of this expectation, agents have learned that maximizing their rewards within the structure of the environment requires acting in accordance with the stereotypic convention; subsequently, the new market decider—learning to predict in an environment with no underlying group differences—nevertheless develops stereotypic expectations for the groups matching those of the previous market decider.

\paragraph{Collective Rewards.} As the market decider in Study 1 learns to represent agents as individuals, the average collective reward obtained by agents increases. We find that in Study 2, the collective reward obtained by agents once the agents and the market decider are fully trained is slightly smaller, but comparable to Study 1 ($B = 0.13$, $SE = 0.02$, $95\% \ CI = [0.08, 0.18]$, $z = 6.21$, $p < .001$, $\text{Cohen's} \ d = 0.35$). This effect is driven by the rewards obtained by the majority agents, who obtain a higher reward than comparable agents in Study 1 ($B = 1.36$, $SE = 0.055$, $95\% \ CI = [1.23, 1.49]$, $z = 24.79$, $p < .001$, $\text{Cohen's} \ d = 0.49$); meanwhile, minority agents obtain a lower reward ($B = -3.62$, $SE = 0.055$, $95\% \ CI = [-3.75, -3.49]$, $z = -66.04$, $p < .001$, $\text{Cohen's} \ d = -1.31$).

In Study 3, the collective reward becomes lower as agents are replaced. By the time all agents and the market decider have been replaced and the market decider is fully trained, the collective reward obtained by the agents is significantly lower than in Study 2 ($B = 1.39$, $SE = 0.01$, $95\% \ CI = [1.37, 1.42]$, $z = 141.18$, $p < .001$, $\text{Cohen's} \ d = 3.65$).

\subsection*{Study 4: Human Trajectories}

As we observed the potential for stereotypes to persist and entrench across generations of artificial agents with no motivations beyond maximizing their own potential returns, we assessed human performance on the same task. If human participants make predictions based on group-based expectations and match others’ expectations through behavioral confirmation, we anticipate that these patterns of behavior will emerge on a task comparable to the simulated environments we developed. Importantly, we anticipate that these behaviors emerge due to the nature of achieving social coordination itself, rather than motivational biases; thus, we expect that the degree of stereotyping shown by participants in these studies will not be correlated to other potential factors associated with stereotyping, such as system justification \citep{jost_decade_2004, jost_role_1994} or social dominance orientation \citep{pratto1994social, sidanius_social_2001}. In other words, the degree to which individuals engage in stereotype-consistent reasoning or behavior in our tasks should not be correlated with the individuals’ motivations to stereotype about social groups.

To test our predictions, we conducted a preregistered study in which 447 human participants completed a web experiment containing an adapted form of the environment from Studies 1–3. In this task, participants were placed in the role of either an agent, collecting resources and attempting to sell them to the market decider, or a market decider, predicting the resources being sold by agents. Participants interacted with a market decider or agents whose behavior trajectories were collected from one run of Study 3, and were randomly presented with a trajectory from the original agents or market, the agents or market after 2 replacements, or the agents and market after all 5 replacements. As most majority agents in Study 3 were predicted to behave skill-consistently, and we wished to compare participants' behavior in the context of being stereotyped or successfully individuated, all participants participating as agents in the task played the role of minority agents. Participants had identical action spaces to the artificial agents, and participants’ skillsets (wood or stone) were counterbalanced between participants. After completing the task, participants completed three social orientation scales: the System Justification Scale \citep[SJS, adapted from][]{vargas2018system}, the Social Dominance Orientation scale \citep[SDO,][]{pratto1994social}, and the Basic Social Justice Orientation scale \citep[BSJO,][]{hulle2018measuring}. Participants in the agent task additionally answered questions about their beliefs regarding the value of different resources, their learning of the task structure, and their performance relative to others. Participants in the market task answered questions about their beliefs regarding the percentage of agents in each group that were skilled at collecting different resources and the fairness of the world they encountered.

\begin{figure*}[!t]
    \centering
    \subfloat[]{\includegraphics[width=.49\linewidth]{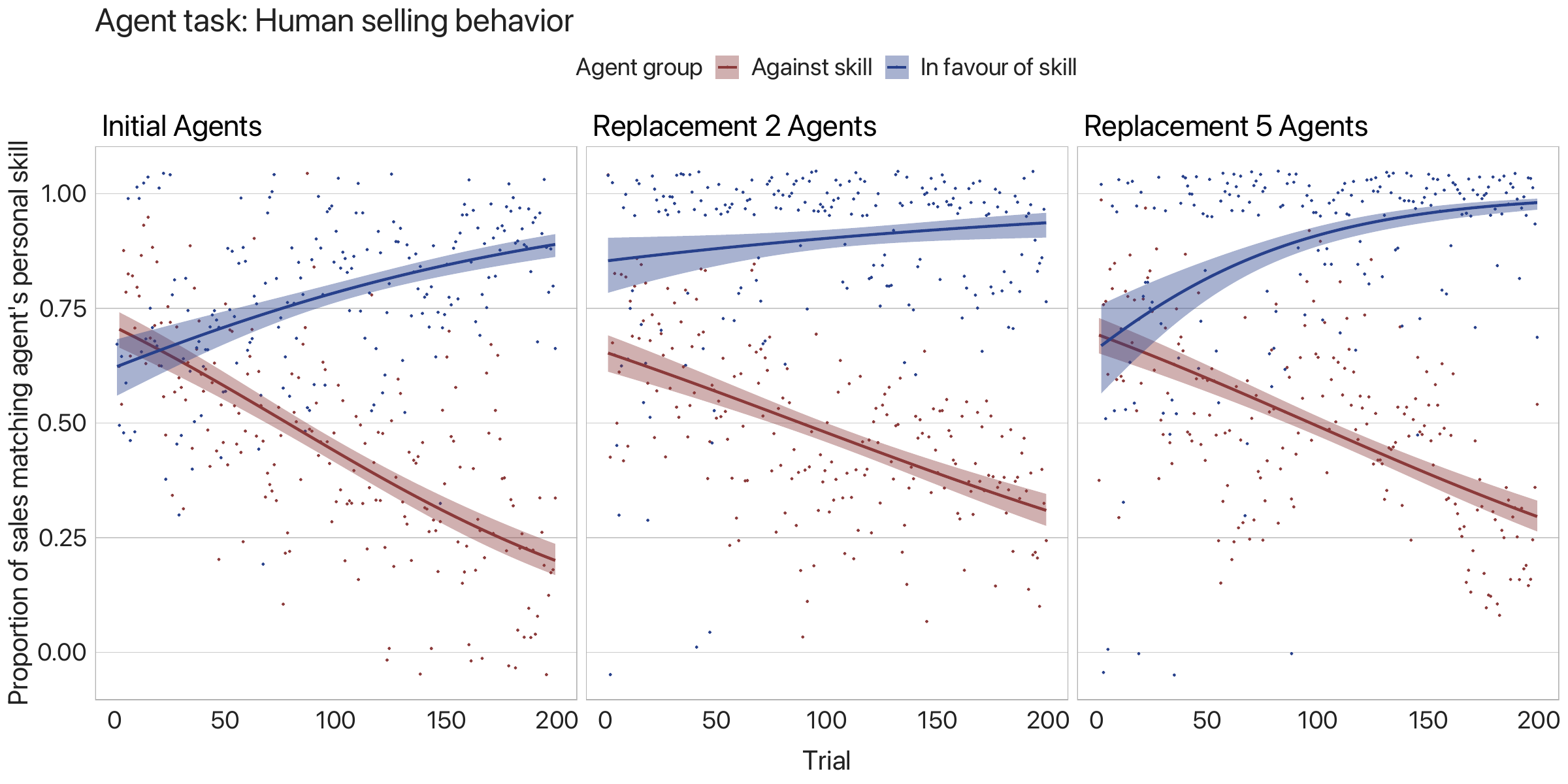}}
    \subfloat[]{\includegraphics[width=.42\linewidth]{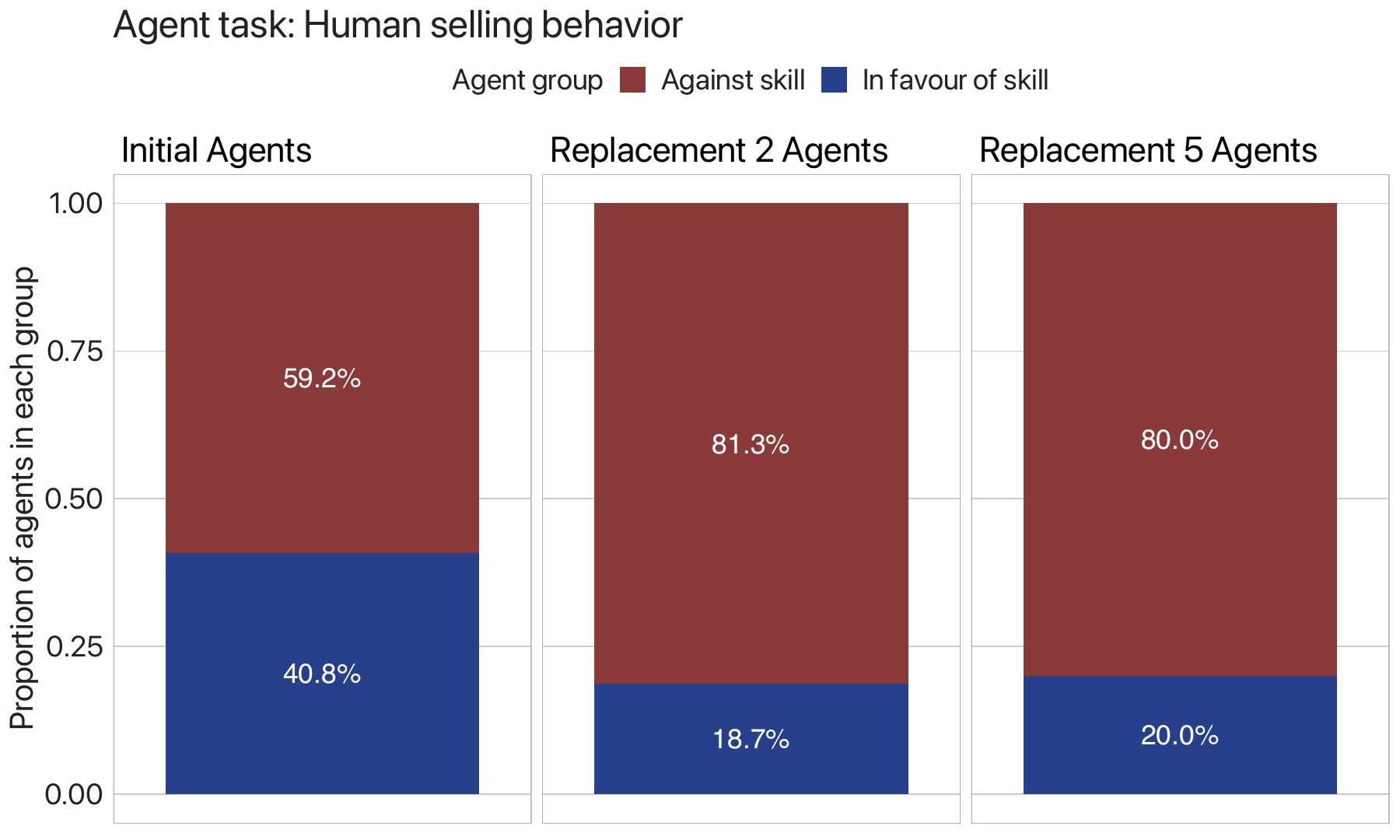}}
    \caption{\textbf{Agent task with human participants.} Participants were assigned to play the role of agents the produce-and-trade task in which the market predicted the agent would bring resources consistent with the agent's true skill (blue) or contrary to the participant's true skill (red). Participants predicted to behave skill-consistently by the market decider progressively made a higher proportion of skill-consistent sales over time, while this proportion declined for participants for which the market made mostly skill-inconsistent prediction (a). Further, the proportion of agents for which the market made skill-consistent predictions was higher for the initial agents than for the agents from Replacement 2 and Replacement 5 (b).}
    \label{fig:fig5}
\end{figure*}

To assess the degree to which human participants demonstrated a similar response pattern to the artificial agents, we first tested human participants’ actions on the agent task. As with the artificial agents, human participants assigned to agents for whom the market decider predicted would bring resources against the participants’ true skill were less likely to bring resources they were skilled at collecting, $B = -2.41$, $SE = 0.171$, $z = -14.12$, $95\% \ CI = [-2.75, -2.08]$, $OR = 0.09$, $p < .001$, and became progressively less likely to do so in later trials, $B = -1.22$, $z = -15.84$, $95\% \ CI = [-1.37, -1.07]$, $p < .001$ (Fig.~\ref{fig:fig5}).

\begin{figure*}[!t]
    \centering
    \includegraphics[width=\textwidth]{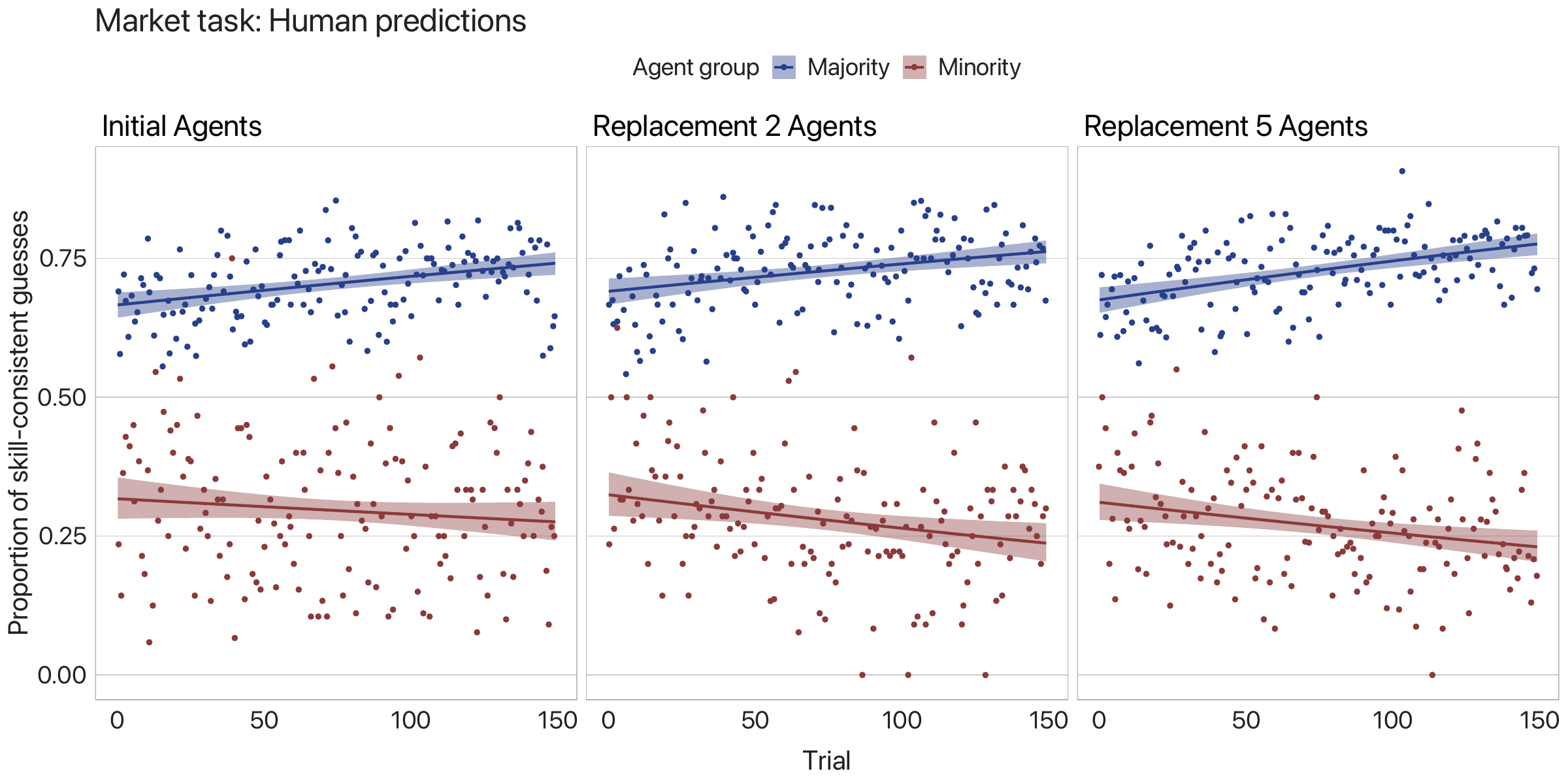}
    \caption{\textbf{Market task with human participants.} Participants were assigned to play the role of the market decider in the produce-and-trade task and predict the resources brought by agents to the market. Across all three conditions, participants predicted that agents who formed the majority within their group would behave more skill-consistently than agents who formed the minority, and this pattern became stronger over time such that in later trials, participants made more stereotypic predictions about both the majority and the minority.}
    \label{fig:fig6}
\end{figure*}

Likewise, human participants on the market decider task displayed a stereotypic effect consistent with the artificial agents. Participants predicted that agents would bring the same resource that a majority of the agents of the same color brought, making them more likely to make skill-consistent predictions for majority agents than for minority agents, $B = 1.97$, $SE = 0.03$, $z = -63.25$, $95\% \ CI = [1.91, 2.03]$, $OR = 7.17$, $p < .001$. This effect also became stronger in later trials, $B = 0.23$, $SE = 0.03$, $z = 7.47$, $95\% \ CI = [0.17, 0.29]$, $OR = 1.26$, $p < .001$ (Fig.~\ref{fig:fig6}).

Participants' beliefs about the degree to which groups of agents shared a single skill (e.g., agents with a purple body color are more likely to attempt to sell stone) also predicted the degree to which they made predictions consistent with that skill for both majority and minority agents. Participants who believed that the majority's skill was more common within a group were more likely to make skill-consistent predictions for majority agents, $B = 0.63$, $SE = 0.026$, $z = 24.46$, $95\% \ CI = [0.58, 0.69]$, $OR = 1.88$, $p < .001$, but were more likely to make skill-inconsistent predictions for minority agents, $B = -0.77$, $SE = 0.035$, $z = 21.98$, $95\% \ CI = [-0.84, -0.70]$, $OR = 0.46$, $p < .001$.

Participants’ differences in SJS and SDO did not meaningfully predict the degree to which they engaged in stereotype-consistent decision-making (SJS: $b = -0.12$, $95\% \ CI = [-0.18, -0.06]$, 100\% in ROPE; SDO: $b = 0.04, 95\% \ CI = [-0.02, 0.10]$, 100\% in ROPE). There was positive evidence that BSJO led participants to less strongly distinguish between the majority and minority agents ($b = 0.27$, $95\% \ CI = [0.20, 0.34]$, 0\% in ROPE).

An analysis of the error rates indicated that being higher in BSJO did not lead participants to be more accurate about minority group members, but rather had higher error rates on majority group members, $B = -0.14$, $SE = 0.04$, $95\% \ CI = [-0.22, -0.06]$, $z = -3.32$, $p < .001$, and equivalent error rates on minority group members, $B = 0.01$, $SE = 0.04$, $95\% \ CI = [-0.07, 0.10]$, $z = 0.33$, $p = .74$. Although this may suggest that BSJO was associated with not using the agents' colour to strategically undermine the structural inequality, this interpretation is less likely given that (a) higher BSJO participants thought that the world was more fair than other participants, $B = 0.22$, $SE = 0.05$, $t(211) = 4.27$, $p < .001$, $\eta^{2}_{p} = 0.08$, and that (b) individual differences in the degree to which these participants saw the game as fair or unfair did not moderate the effect of BSJO on error rates ($B = -0.24$, $SE = 1.40$, $t(209) = -0.17$, $p = .86$, $\eta^{2}_{p} < 0.001$).

\section*{Discussion}

In this study, we demonstrate that agents rapidly regularize around group-based social conventions that facilitate a simple coordination strategy. This tendency was particularly strong in larger populations, in which learning the skills of each individual was more time-consuming and costly, initially providing a predictive advantage to a convention-based strategy. However, as this regularity became established, it became a learning signal in its own right, making it difficult to dislodge the convention for both agents and market deciders, as well as the human participants playing these roles.

Human participants behaved comparably to the predictions made by the models, acting in accordance with the market decider’s predictions even when the prediction was that the participant would bring a resource they were unskilled in collecting, and developing stereotypic expectations about agents that were based on their visual appearance rather than their individual skills. Critically, the associations between adopting these stereotypic conventions and achieving higher rewards were not simply adopted by the human participants as a useful heuristic for individuals to maximize their own expected utility irrespective of the underlying structure of the world, but were seen as a reflection of the true state of the world.

As participants considered the strong learning signal of the groups' distinct behavior to reflect real essential differences between groups, participants' choices were not affected by their attitudes towards real-world social inequality, that might otherwise have provided a motivation for individuals to behave stereotypically even though doing so could disadvantage minority members of a group. In fact, participants with higher scores in BSJO believed the structure of the world within our tasks was more fair than those with lower BSJO scores.

These findings suggest that stereotyping may be particularly prone to emerging—and particularly resistant to change—in larger populations. In these contexts, one's experience with a single individual of a group is a small part of one's experience with the group as a whole. As a result, individuals within these societies may be more likely to perceive the differences they observe to reflect real underlying differences rather than an artifact of the demands of social coordination. Modern social groupings, with sizes in the millions and more, may be particularly vulnerable to the development and persistence of stereotypes, posing a unique threat to the potential for diversity and individuality within contemporary societies. Because these expectations dictate the structure of the subsequent coordination, expectations can amplify stereotypic behaviors and lead them to persist even if the factors that initially differetiated groups disappear. 

As a result of these stereotypic conventions, agents in these environments with skills that do not match the stereotype are presented with a double bind: if they continue to attempt to sell the resource they are skilled at collecting, they will often fail to coordinate with the market decider or obtain any rewards. However, if they sell the resource they are unskilled at collecting, they strengthen the market decider's subsequent expectations. This dynamic leads agents to suppress their own individual differences, acting in accordance with the stereotype even when their skills are unsuited to it.

The fact that human participants on our tasks made similar choices despite differing degrees of motivation to reject social inequality suggests that these humans' egalitarian attitudes and aversion to stereotyping were unable to prevent the emergence of similar degrees of unequal outcomes for the artificial agents, because most participants believed that the differences in the behavior they observed were reflective of a fair society in which groups are largely homogeneous and differ based on their skills, not simply because human beings were maximizing their own rewards.

Although the threat of stereotyping is often recognized as a social problem in that people identify and endorse moral norms against the use of stereotypes \citep{cao_people_2019, mae_hoist_2005}, including across political ideologies \citep{vrantsidis_stereotypes_nodate}, our results may shed light on why stereotyping remains a ubiquitous feature of human societies. If behavioral confirmation can generate and perpetuate stereotypic norms about how members of a group ought to behave, then subsequent generations observe real-world confirmation of these stereotypic beliefs, making it seem as if the expectation was justified all along. Similarly, human participants on our tasks engaged in stereotypic behavior, but they did so because they believed that their beliefs about the agents were statistically accurate generalizations.

The fact that patterns of stereotypic behavior can remain in place even when the underlying group differences are no longer consistent with the stereotyping should serve as a caution against succumbing to a “naturalistic fallacy” in stereotyping: the idea that observed differences between groups is due to an underlying or essential characteristic of the groups, rather than due to other explanations such as structural or historical factors. For example, the existence of the “gender-equality paradox”—the observation that gender segregation by occupation is stronger in gender-egalitarian countries—has sometimes been explained as revealing intrinsic differences between men's and women's interests when given a free choice to pursue a career \citep[e.g.,][]{stoet2018gender}. However, consistent with stereotype-based explanations for the gender-equality paradox \citep[e.g.][]{breda2020gender}), our data illustrate how beliefs about what is perceived to be a rational decision process driven by free choice, such as choosing one's career, could become biased by the development of these stereotypic norms.

The mechanisms by which perpetuation of stereotypic conventions can occur also provides us a method for understanding how inequalities persist. For example, although the coordination tasks we introduce in these studies are in principle equitable with regards to the rewards provided to agents, in practice a proportion of agents were disadvantaged by the stereotype, because they were not able to live up to their full potential. Furthermore, many real-world instances of coordination result in divisions of labour that are unequal, particularly to structurally disadvantaged or numerically smaller groups \citep[e.g.][]{bruner_minority_2019, oconnor_emergence_2019}). Thus, the effects we observe in this work can be extended to understand when conformity to a stereotypic convention may emerge even when a group as a whole is disadvantaged by the convention, rather than disadvantaging a minority of agents within a group.

\section*{Materials and Methods}

\subsection*{Environment Setup}

The task for our agents $i \in I = \{1 ... N\}$ can be formalized as a partially-observable Markov decision process (POMDP). Each agent $i$ has an observation size $v$, observation function $\mathcal{O}_i : \mathcal{S} \rightarrow \mathbb{R}^v$, and action set $\mathcal{A}_i$; for the producing agents, $v = 6$ and $\mathcal{A}_i = 7$, and for the market decider, $v = 16$ and $\mathcal{A}_i = 2$. State transitions are determined according to the transition probability $\mathcal{T}(\mathcal{S}'|\mathcal{S}, \mathcal{A})$, with each agent receiving rewards $r_i \in \mathcal{R}$ according to the reward function $\mathcal{R}: \mathcal{S} \times \mathcal{A} \rightarrow \mathbb{R}$. 

At each timestep, the producing agents observe a 6-digit vector indicating their own resource pool (quantity of coins, stone, and wood, and whether each respective quantity is above 0), and select an action from a list of 7 possible actions (extract stone, extract wood, build, sell stone, sell wood, buy stone, buy wood). Simultaneously, the market decider observes the 16-digit binary vector representing the identity of the agent that attempts to sell stone or wood, and selects one of two actions (predict the agent will sell stone or wood). Both the agents and the market decider must learn a policy mapping observed states to action probabilities based on the long-term expected discounted reward, $E_i \Bigl[ \sum_{t=0}^{\infty} \gamma^{t} r_{i,t} \Bigr]$.

In all studies, the probability that certain actions (extracting stone, extracting wood, extracting skill) will be successfully completed by agents is dependent on their skill level. Agents are instantiated with one of three specializations: chopping, mining, or building specialist. The choppers are good at chopping (0.75) and bad at mining (0.25) and building (0.05). The miners are good at mining (0.75) and bad at chopping (0.25) and building (0.05). The builders are good at building (0.95) and bad at chopping and mining (both 0.1). Agents receive 15 points of reward for successfully building a house, receive 1 point of reward for successfully selling wood or stone, and lose 2 points of reward for buying wood or stone from the market decider.

\subsection*{Agent Architecture and Training Method}

In all three studies, each agent and market decider comprised an independent neural network (including the agents serving as replacers in Study 3). These networks were identical in architecture and were initialized with random weights at the start of each experimental condition. All models were structured using an actor-critic architecture \citep{haarnoja_soft_2018}. Each model consisted of an actor network producing an action policy and a critic network outputting a state value estimate. Both networks consisted of a three-layer fully connected multilayer perceptrons (MLP) (Actor and Critic) with 64 hyperbolic tangent units (tanh) in the first and third layers and 128 units in the second layer. The critic network outputs a scalar representing the estimated value of the current state. The actor network produces a probability over the actions to be chosen (7 actions for the agents and 2 for the market). 

What the agents and the market decider observe shares the same structure across the three studies. At each step within an epoch, each agent observed the state of its current possessions, represented by a 6-digit vector: the quantity of stones, the quantity of wood, the quantity of coins, whether the number of wood is greater than 1, whether the number of stones is greater than 1, and whether the number of coins is larger than 1. If any agent decides to sell resources to the market, the market observes a 16-digit binary code representing the identity of that agent. After transformations of the input observations, the agent’s and the market’s neural networks generate an action policy and an estimate of the current state. The observed state, the action policy, and the reward outcome of each time step are stored for training the neural networks. Each agent and market decider have a separate rollout buffer to store the experience trajectories. 
Specifically, we train the neural networks using the proximal policy optimization algorithm \citep[PPO,][]{schulman_proximal_2017}, with an Adam optimizer \citep{kingma_adam_2014}. The learning rates are $10^{-3}$ and $5 \times 10^{-4}$ for the actor network and the critic network respectively, and the discount factor $\gamma$ is 0.9. At the end of each epoch, all agents and the market are trained with their own stored experiences. In detail, the stored trajectory of experiences in the current epoch are retrieved from the memory buffer, and the agents/market maximize the following objective:

\begin{equation}
    L_t^{\text{CLIP}+VF+S}(\theta) = \mathbb{\hat{E}}_t[L_t^{\text{CLIP}}({\theta}) - c_1 L_t^{VF}({\theta}) + c_2 S[\pi_{\theta}](s_t)],
\end{equation}

\begin{equation}
    L_t^{\text{CLIP}}({\theta}) = \mathbb{\hat{E}}_t[\text{min}(r_t ({\theta}) \hat{A_t}, \text{clip}(r_t ({\theta}), 1-{\epsilon}, 1+{\epsilon}) \hat{A_t})]
\end{equation}

where $c_1$, $c_2$ are coefficients, and $S$ denotes an entropy bonus, and $L_t^{VF}$ is a squared-error loss $(V_{\theta}(s_t) - V_t^{\text{targ}})^2$. $V_{\theta}(s_t)$ denotes the generated estimate of the value of the current state, and $V_t^{\text{targ}}$ denotes the actual value of the current state. $\hat{A_t}$ denotes an estimator of the advantage function at timestep $t$. The term $r_t ({\theta})$ denotes the probability ratio between the current stochastic policy and the old stochastic policy with which an agent collected the experience to learn from. The function $\text{clip}()$ establishes a bound for the probability ratio term $r_t ({\theta})$ within the interval $[1-{\epsilon}, 1+{\epsilon}]$. In our study, $c_1 = 0.5$, $c_2 = 0.01$, ${\epsilon} = 0.2$.

\subsection*{Experimental Design: Human Task}

\subsubsection*{Participants and Design} 447 adults (Sex: Male = 281, Female = 165, Not provided = 1. Age: $M$ = 38.34, $SD$ = 12.65, Range = [18, 74]. Nationality: United Kingdom = 184, United States = 117, Canada = 71, Australia = 33, Nigeria = 9, New Zealand = 8, South Africa = 4, Ireland = 4, India = 3, Poland = 2, Others = 12. Ethnicity: White = 321, Asian = 53, Black = 33, Mixed = 29, Other = 7, Not provided = 5) were recruited from the online platform Prolific. To be eligible, participants had to be native speakers of English and have an acceptance rate of at least 99\% on the platform to qualify. Participants were paid £2.25 for completing the task, and a bonus of £0.25 for scoring in the top 10\% of participants. This study was approved by the University of Toronto Research Ethics Board (\#33582).

Participants were randomly assigned to complete either the market decider task ($N$ = 226) or the agent task ($N$ = 221) and were also randomly assigned to complete the task at either the early timepoint, before any agents were replaced ($N$ = 149), at the middle time point, after 40\% of the original agents were replaced ($N$ = 144), or at the late time point, after 100\% of the original agents were replaced ($N$ = 154). As these conditions were completed between subjects, no participant performed both tasks or encountered the task at more than one time point.

According to our preregistered inclusion criteria, participants were excluded from analysis if their responses were multivariate outliers on Mahalanobis distance with an $\alpha$ of .001. 10 participants’ responses were excluded from the agent task and 13 were excluded from the market task, leaving a total of 424 final participants included in the analysis.

\subsubsection*{Materials and Procedure}

After reading and agreeing to the consent form, participants completed either the market game or the agent game in a ReactJS web app.

In both tasks, the market behavior (predicting wood or stone for each agent approach) and the agent behavior (performing an action such as collecting, buying, or selling a resource or building a house) was drawn from one run of Study 3. These behaviors were transformed into action probabilities for the market or agents that the human participants encountered in Study 4.

\paragraph{Market Game.} Participants playing the market game were first presented with instructions on how to play the game. They were told that they were going to encounter a number of aliens, whose identity could be determined by their body color and unique 12-pixel nametag. Participants then needed to predict the resources that each alien would bring to the market. They were told that successfully predicting an alien approach would gain them points, and that the goal in the task was to maximize the number of points obtained.

Then, participants completed 150 total trials of the game. On each trial, one alien would be shown and the participant selected whether they thought the alien brought wood or stone to the market. The individual alien that was shown on a given trial was determined according to the approach probability (how often each agent came to the market) and the resource probability (how often each agent brought wood or stone to the market) of agents in one run of Study 3. The agent body color (purple, yellow, or cyan) was determined by the first three digits of the agents’ 16-digit code identifier, and each agent was assigned a unique 12-digit identifier in place of the remaining 13 digits. As there were a total of 300 agents, not all identifiers were used.

Thus, agents who approached the market more often were more likely to be shown to a participant, and agents with a higher wood probability were more likely to bring wood than stone to the market.

After completing the task, participants were presented with a short survey. They were asked about their beliefs about the group’s general skill at collecting wood or stone, the percentage of each group that was better at collecting wood or stone, and their beliefs about the fairness of the task, as well as three social orientation scales: the System Justification Scale (SJS), the Social Dominance Orientation scale (SDO), and the Basic Social Justice Orientation scale (BSJO).

\paragraph{Agent Game.} Participants playing the agent game were first presented with instructions on how to play the game. They were told that they were a worker trying to obtain coins, which could be obtained by selling resources or building houses, and that each worker has a skill that makes them better at collecting different kinds of resources. Further, they were told that in order to successfully sell a resource, they needed to bring the same resource that the market decider predicted they would bring.

Participants’ skill was randomly counterbalanced between being skilled at collecting wood or stone. No participants were skilled at building houses. Participants’ chance of successfully collecting a resource depended on their skill, with a 75\% chance to succeed at collecting their skilled resource and a 25\% chance to succeed at collecting their unskilled resource. All participants had a 5\% chance of successfully building a house.

At the beginning of the task, participants were assigned an agent identity code that was not visible to the participant. The market decider’s predictions for the participant were computed for each epoch (from 1 to 200) as the average proportion of predictions that the agent would bring the resource they were skilled at collecting. Thus, the markets’ prediction for agents on each trial would vary from 0 (always predict against the agents’ true skill) to 1 (always predict consistently with the agents’ true skill).

Participants completed 200 trials of the task, corresponding to 1 epoch of the task. Similar to the agents in Studies 1–3, participants could attempt to collect wood, collect stone, buy wood, buy stone, sell wood, sell stone, or build a house. Participants received 1 coin for successfully selling a resource, and 15 coins for successfully building a house.

If participants attempted to sell to the market on a given trial, the market would make a prediction for the agent based on the probability that the agent would bring their skilled resource in the corresponding epoch.

After completing the task, participants were presented with a short survey. They were asked whether they believed the market needed more wood or stone, whether wood or stone was more valuable to own, how well they learned the structure of their environment, how well they believed they did relative to other players, and how frustrating and enjoyable they found the task. They were also presented with three social orientation scales (SJS, SDO, BSJO).

\subsection*{Statistical Analyses}

Prior to analysis, the following variables of interest were collected or computed for Studies 1–3. For the market, we used the proportion of the market's predictions that were consistent with the agent's skills for a given epoch and run. For each agent, we computed the proportion that the agent attempted to sell the resource it was skilled at collecting relative to the total number of sales attempted (skilled and unskilled), as well as the proportion of skilled extractions relative to the total number of extractions (skilled and unskilled) for a given epoch and run.

For the three simulation studies, we used linear mixed-effects regression (LMER) models which predict the degree of skill-based predictions by the market, or the degree of skilled extraction or skilled sales by agents. For all models, we first compute an omnibus ANOVA and compute $F$-test and partial $\eta^{2}$ values for the predictors of interest, and then compute follow-up pairwise contrasts adjusted using Tukey's method for multiple comparisons. The main analyses in Studies 1–3 used the following predictors:

\begin{equation}
    \text{Ratio} = 1 + \text{epoch} * \text{population\_size} + (1|\text{Simulation}) \ \text{(Study 1)}
\end{equation}
\begin{equation}
    \text{Ratio} = 1 + \text{epoch} * \text{population\_size} * \text{group} + (1|\text{Simulation}) \ \text{(Study 2)} 
\end{equation}
\begin{equation}
    \text{Ratio} = 1 + \text{epoch} * \text{group} * \text{replacement} + (1|\text{Simulation}) \ \text{(Study 3)}
\end{equation}

\noindent where $\text{Ratio}$ is the proportion of extraction, sales, and market prediction that are consistent with an agent's true underlying skill. Although population size is a numerical quantity, we analyze it as a factor as the values chosen are four discrete quantities and may display non-linear effects at different population sizes. The values for epoch are scaled and normalized such that the zero point is placed at epoch 100. Note that in Study 2, the group variable included the data from Study 1 as a separate group, in order to allow us to compare data from the majority and minority in Study 2 to a non-stereotyped baseline.

Additionally, we recorded the average reward obtained per group (Study 1: all agents, Studies 2 and 3: mining and chopping specialist majorities and minorities) throughout the three studies. We then compute the population-level average reward obtained per run by the total population. These outcomes were compared using LMER models, with follow-up pairwise contrasts and effect sizes computed between groups (Study 1 Agents, Study 2 Majority, Study 2 Minority) and between study populations as a whole (Study 1, Study 2, Study 3).

In Study 4, we use a similar pipeline for assessing the data as in Study 3. Due to a data collection error, actions were not recorded on the final trial for 100 out of the 221 participants in the agent task. The first 199 trials were not affected for any participants. Consistent with our preregistration, we used Mahalanobis distance to identify multivariate outliers with $\alpha < .001$, indicating careless or repetitive responses on the task or survey. 11 participants were excluded from the agent task and 13 participants were excluded from the market task, yielding a final sample of 210 participants on the agent task and 213 participants on the market task.

As participants performed one action per trial instead of multiple actions in one epoch (as in the simulation studies), we use a generalized linear mixed-effects regression (GLMER) with a binary outcome to analyze participants' actions, with effect sizes measured as odds ratios ($OR$) and follow-up pairwise contrasts computed for measures of interest. For the agent task, we focus on whether the participant's extracting or selling action was the skilled (1) or unskilled (0) resource. For the market task, predictions are either skill-consistent (1) or skill-inconsistent (0). For SJS, SDO, and BSJO scores, we compute and normalize participants' full scale scores before analysis. In Study 4, we test the following analyses:

\begin{equation}
    \begin{aligned}
        \text{Ratio} &= 1 + \text{epoch} * \text{group} + \text{replacement} + \\
        \text{SJS} &+ \text{group:SJS} + (1|\text{Participant})
    \end{aligned}  
\end{equation}
\begin{equation}
    \begin{aligned}
        \text{Ratio} &= 1 + \text{epoch} * \text{group} + \text{replacement} + \\
        \text{SDO} &+ \text{group:SDO} + (1|\text{Participant})
    \end{aligned} 
\end{equation}
\begin{equation}
    \begin{aligned}
        \text{Ratio} &= 1 + \text{epoch} * \text{group} + \text{replacement} + \\
        \text{BSJO} &+ \text{group:BSJO} + (1|\text{Participant})
    \end{aligned}
\end{equation}

For our preregistered predictions of a null effect on SJS, SDO, and BSJO, we used a Bayesian GLMER using a region of practical equivalence (ROPE) test, using a prior of $b \sim \mathcal{N}(0,5)$ on all parameters. For our models, we follow Kruschke \citep{kruschke2018rejecting} and interpret a negligible effect as $0.1 \times SD_{y}$, where the standard difference of a parameter expressed in the log odds scale is $\pi / \sqrt{3}$, yielding a ROPE of $[-0.18, 0.18]$.

\section*{Acknowledgments}
The authors wish to thank Gillian Hadfield, Raphael Köster, Joel Z. Leibo, Paul Stillman, and the members of the Social AI Lab at the University of Toronto for their helpful feedback.

\section*{Data availability}
\raggedright{The code used for the multi-agent environment, agent architectures, and model training is available at: \faGithub $ \ $\href{https://github.com/Yikai369/Stereotypic-expectations-entrench-unequal-conventions-across-generations}{github.com/Yikai369}}

\nolinenumbers

\bibliography{library}

\bibliographystyle{unsrtnat}

\end{document}